\documentclass[showpacs,oneside,superscriptaddress,pra,twocolumn]{revtex4}%
\usepackage{amsmath}
\usepackage{amsfonts}
\usepackage{amssymb}
\usepackage{graphicx}%
\begin{document}
\title{Relativistic-invariant quantum entanglement between the spins of moving bodies}
\author{Hui Li}
\email{lhuy@mail.ustc.edu.cn}
\affiliation{Department of Modern Physics, University of Science and Technology of China,
Hefei, 230027, People's Republic of China}
\author{Jiangfeng Du}
\email{djf@ustc.edu.cn}
\affiliation{Department of Modern Physics, University of Science and Technology of China,
Hefei, 230027, People's Republic of China}
\affiliation{Department of Physics, National University of Singapore, Lower Fent Ridge,
Singapore 119260, Singapore}

\begin{abstract}
The entanglement between spins of a pair of particles may change because the
spin and momentum become mixed when viewed by a moving observer [R.M. Gingrich
and C. Adami, Phys. Rev. Lett. \textbf{89}, 270402 (2002)]. In this paper, it
is shown that, if the momenta are appropriately entangled, the entanglement
between the spins of the Bell states can remain maximal when viewed by any
moving observer. Further, we suggest a relativistic-invariant protocol for
quantum communication, with which the non-relativistic quantum information
theory could be invariantly applied to relativistic situations.

\end{abstract}
\maketitle

\section{Introduction}

The relativistic thermodynamics has been an intriguing problem for decades
\cite{0}. It has been shown that probability distributions can depend on the
frames, and thus the entropy and information may change if viewed from
different frames \cite{a7}. Recently, the effect of Lorentz boosts on quantum
states, quantum entanglement and quantum information has drawn particular
interests \cite{1,2,5,6}. The relativistic quantum information theory may
become necessary in the near future, with possible applications to quantum
clock synchronization \cite{a5} and quantum-enhanced global positioning
\cite{a6}.

Entanglement of quantum systems forms a vital resource for many quantum
information processing protocols \cite{a1}, including quantum teleportation
\cite{a2}, cryptography \cite{a3} and computation \cite{a4}. However, it has
shown that Lorentz boosts can affect the marginal entropy of a single quantum
spin \cite{1}, and a fully entangled spin-1/2 system may lose entanglement
when observed by a Lorentz-boosted observer \cite{2}. Particularly, fully
entangled spin states in the rest frame will most likely decohere due to the
mixing with momentum if viewed from a moving frame, depending on the initial
momentum wave function \cite{2}. The entanglement between two systems depends
on the frame in which this entanglement is measured. These effects may have
important consequences for quantum communication, especially when the
communicating parties are in relative movement.

In this paper, we show that for a pair of spin-1/2 massive particles, if the
momenta are appropriately entangled, the entanglement between the spins can
remain the same as in the rest frame when viewed from any Lorentz-transformed
frame. We also find a set of states, of which the marginal entropy,
entanglement and measurement results of the spins are independent of the
frames from which they are observed. Further, we suggest a
relativistic-invariant representation of the quantum bit (qubit), and suggest
a relativistic-invariant protocol for quantum communication, with which the
non-relativistic quantum information theory could be invariantly applied to
relativistic situations. Though, in this paper, we restrict ourselves to
spin-1/2 cases, the generalization to larger spins could be done analogously.
Particularly, the generalization to spin-1 massless particles, such as photons
\cite{6}, may be of special interests since current experiments for quantum
communications are mostly based on photons.

\section{Entanglement between the spins, with the presence of momentum
entanglement}

We start by investigating the bipartite state that, in the momentum
representation, has the following form viewed from the rest frame,%
\begin{equation}
\Psi\left(  \mathbf{p},\mathbf{q}\right)  =g\left(  \mathbf{p},\mathbf{q}%
\right)  \left\vert \psi^{-}\right\rangle ,\label{eq 1}%
\end{equation}
where $\mathbf{p}$ and $\mathbf{q}$ are the momenta for the first and second
particles, respectively (For review of the definition of the momentum
eigenstates for massive particles with spin and the transformation under
Lorentz boosts, one may refer to Refs. \cite{3,1,2}). The spin part of the
state is the singlet Bell state%
\begin{equation}
\left\vert \psi^{-}\right\rangle =\frac{1}{\sqrt{2}}\left(  \left\vert
\uparrow\downarrow\right\rangle -\left\vert \downarrow\uparrow\right\rangle
\right)  ,\label{eq 2}%
\end{equation}
where $\left\vert \uparrow\downarrow\right\rangle =\left\vert \uparrow
\right\rangle \otimes\left\vert \downarrow\right\rangle $, $\left\vert
\downarrow\uparrow\right\rangle =\left\vert \downarrow\right\rangle
\otimes\left\vert \uparrow\right\rangle $, with%
\begin{equation}
\left\vert \uparrow\right\rangle =\left(
\begin{array}
[c]{c}%
1\\
0
\end{array}
\right)  ,\left\vert \downarrow\right\rangle =\left(
\begin{array}
[c]{c}%
0\\
1
\end{array}
\right)  .
\end{equation}
The momentum distribution $g\left(  \mathbf{p},\mathbf{q}\right)  $ is
normalized according to%
\begin{equation}
\iint\left\vert g\left(  \mathbf{p},\mathbf{q}\right)  \right\vert
^{2}\widetilde{\text{d}}\mathbf{p}\widetilde{\text{d}}\mathbf{q}%
=1,\label{eq 22}%
\end{equation}
where $\widetilde{\text{d}}\mathbf{p}$ ($\widetilde{\text{d}}\mathbf{q}$) is
the Lorentz-invariant momentum integration measures given by (We use natural
units: $c=1$.)
\begin{equation}
\widetilde{\text{d}}\mathbf{p=}\frac{\text{d}^{3}\mathbf{p}}{2\sqrt
{\mathbf{p}^{2}+m^{2}}}.\label{eq 23}%
\end{equation}
Note that there is no entanglement between the spin and the momentum parts of
$\Psi\left(  \mathbf{p},\mathbf{q}\right)  $. The spins are maximally
entangled, while the entanglement between momenta depends on $g\left(
\mathbf{p},\mathbf{q}\right)  $. In what follows, we use $\mathbf{p}$ to
represent the momentum $4$-vector as in Eq. (\ref{eq 5}) unless it is ambiguous.

To an observer in a frame Lorentz transformed by $\Lambda^{-1}$, the state
$\Psi\left(  \mathbf{p},\mathbf{q}\right)  $ appears to be transformed by
$\Lambda\otimes\Lambda$. Therefore the state viewed by this observer appears
to be%
\begin{align}
\Psi^{\prime}\left(  \mathbf{p},\mathbf{q}\right)   &  =U\left(
\Lambda\otimes\Lambda\right)  \Psi\left(  \mathbf{p},\mathbf{q}\right)
\nonumber\\
&  =\left[  U_{\Lambda^{-1}\mathbf{p}}\otimes U_{\Lambda^{-1}\mathbf{q}%
}\right]  \Psi\left(  \Lambda^{-1}\mathbf{p},\Lambda^{-1}\mathbf{q}\right)
,\label{eq 20}%
\end{align}
where $U\left(  \Lambda\otimes\Lambda\right)  $ represents the unitary
transformation induced by the Lorentz transformation. Here, for compactness of
notation, we define $U_{\mathbf{p}}\equiv D^{\left(  1/2\right)  }\left(
R\left(  \Lambda,\mathbf{p}\right)  \right)  $ as the spin-1/2 representation
of the Wigner rotation $R\left(  \Lambda,\mathbf{p}\right)  $ \cite{2,3}.
Because $\Psi^{\prime}\left(  \mathbf{p},\mathbf{q}\right)  $ differs from
$\Psi\left(  \mathbf{p},\mathbf{q}\right)  $ by only local unitary
transformations, the entanglement will not change provided we do not trace out
a part of the state. However, in looking at the entanglement between the
spins, tracing out over the momentum degrees of freedom is implied. In
$\Psi^{\prime}\left(  \mathbf{p},\mathbf{q}\right)  $ the spin and momentum
may appear to be entangled, therefore the entanglement between the spins may
change when viewed by the Lorentz-transformed observer. By writing
$\Psi^{\prime}\left(  \mathbf{p},\mathbf{q}\right)  $ as a density matrix and
tracing over the momentum degrees of freedom, the entanglement between the
spins (viewed by the Lorentz-transformed observer) could be obtained by
calculating the Wootters' concurrence \cite{4} of the reduced density matrix
for spins.

Any Lorentz transformation could be written as a rotation followed by a boost
\cite{3}, and tracing over the momentum after a rotation will not change the
spin concurrence \cite{2}, therefore we can look only at pure boosts. Without
loss of generality we may choose boosts in the $z$-direction and write the
momentum 4-vector in polar coordinates as%
\begin{equation}
\mathbf{p=}\left(  E_{\mathbf{p}},p\cos\varphi_{\mathbf{p}}\sin\theta
_{\mathbf{p}},p\sin\varphi_{\mathbf{p}}\sin\theta_{\mathbf{p}},p\cos
\theta_{\mathbf{p}}\right)  ,\label{eq 5}%
\end{equation}
with $E_{\mathbf{p}}=\sqrt{p^{2}+m^{2}}$, $0\leqslant\theta_{\mathbf{p}%
}\leqslant\pi$ and $0\leqslant\varphi_{\mathbf{p}}<2\pi$. Let $\Lambda\equiv
L\left(  \mathbf{\xi}\right)  $ be the boost along the $z$-direction (as
defined in Ref. \cite{2}), where $\mathbf{\xi}$ is the rapidity of the boost
and let $\xi=\left\vert \mathbf{\xi}\right\vert $. With Eq. (\ref{eq 5}), we
obtain%
\begin{equation}
U_{\mathbf{p}}=\left(
\begin{array}
[c]{cc}%
\alpha_{\mathbf{p}} & \beta_{\mathbf{p}}e^{-i\varphi_{\mathbf{p}}}\\
-\beta_{\mathbf{p}}e^{i\varphi_{\mathbf{p}}} & \alpha_{\mathbf{p}}%
\end{array}
\right)  ,\label{eq 6}%
\end{equation}
where%
\begin{align}
\alpha_{\mathbf{p}} &  =\sqrt{\frac{E_{\mathbf{p}}+m}{E_{\mathbf{p}}^{\prime
}+m}}\left(  \cosh\frac{\xi}{2}+\frac{p\cos\theta_{\mathbf{p}}}{E_{\mathbf{p}%
}+m}\sinh\frac{\xi}{2}\right)  ,\\
\beta_{\mathbf{p}} &  =\frac{p\sin\theta_{\mathbf{p}}}{\sqrt{\left(
E_{\mathbf{p}}+m\right)  \left(  E_{\mathbf{p}}^{\prime}+m\right)  }}%
\sinh\frac{\xi}{2},\label{eq 7}%
\end{align}
and $E_{\mathbf{p}}^{\prime}=E_{\mathbf{p}}\cosh\xi+p\cos\theta_{\mathbf{p}%
}\sinh\xi$. The similar is for the second particle with momentum $\mathbf{q}$.
Substituting Eq. (\ref{eq 6}) into Eq. (\ref{eq 20}), we obtain the state
viewed by the Lorentz-boosted observer as%
\begin{equation}
\Psi^{\prime}\left(  \Lambda\mathbf{p},\Lambda\mathbf{q}\right)
=\frac{g\left(  \mathbf{p},\mathbf{q}\right)  }{\sqrt{2}}\left(
\begin{array}
[c]{c}%
\alpha_{\mathbf{p}}\beta_{\mathbf{q}}e^{-i\varphi_{\mathbf{q}}}-\alpha
_{\mathbf{q}}\beta_{\mathbf{p}}e^{-i\varphi_{\mathbf{p}}}\\
\alpha_{\mathbf{p}}\alpha_{\mathbf{q}}+\beta_{\mathbf{p}}\beta_{\mathbf{q}%
}e^{-i\left(  \varphi_{\mathbf{p}}-\varphi_{\mathbf{q}}\right)  }\\
-\alpha_{\mathbf{p}}\alpha_{\mathbf{q}}-\beta_{\mathbf{p}}\beta_{\mathbf{q}%
}e^{i\left(  \varphi_{\mathbf{p}}-\varphi_{\mathbf{q}}\right)  }\\
\alpha_{\mathbf{p}}\beta_{\mathbf{q}}e^{i\varphi_{\mathbf{q}}}-\alpha
_{\mathbf{q}}\beta_{\mathbf{p}}e^{i\varphi_{\mathbf{p}}}%
\end{array}
\right)  .\label{eq 8}%
\end{equation}

At the present stage, we use an \textquotedblleft entangled
Gaussian\textquotedblright\ with width $\sigma$ for the momentum distribution,
as follows,%
\begin{align}
&  g\left(  \mathbf{p},\mathbf{q}\right)  \nonumber\\
&  =\sqrt{\frac{1}{N}\exp\left[  -\frac{\mathbf{p}^{2}+\mathbf{q}^{2}}%
{4\sigma^{2}}\right]  \exp\left[  -\frac{\mathbf{p}^{2}+\mathbf{q}%
^{2}-2x\mathbf{p}\cdot\mathbf{q}}{4\sigma^{2}\left(  1-x^{2}\right)  }\right]
},\label{eq 3}%
\end{align}
where $x\in\left[  0,1\right)  $ and $N$ is the normalization. In Eq.
(\ref{eq 3}), for a given $\sigma$, $x$ could be reasonably regard as a
measure of the entanglement between momenta. When $x=0$, the momentum part of
the state is separable, \textit{i.e.} the momentum entanglement is zero.
However at the limit $x\rightarrow1$, we have%
\begin{equation}
\lim_{x\rightarrow1}g\left(  \mathbf{p},\mathbf{q}\right)  =\sqrt{\frac
{1}{N^{\prime}}\exp\left[  -\frac{\mathbf{p}^{2}}{2\sigma^{2}}\right]
\delta^{3}\left(  \mathbf{p}-\mathbf{q}\right)  },\label{eq 4}%
\end{equation}
where $N^{\prime}$ is the normalization. Eq. (\ref{eq 4}) indicates a perfect
correlation between the momenta. Note that in Eq. (\ref{eq 4}) the momenta are
not necessarily maximally entangled.

\begin{figure}
[t]
\begin{center}
\includegraphics[
height=5.8474cm,
width=7.5981cm
]%
{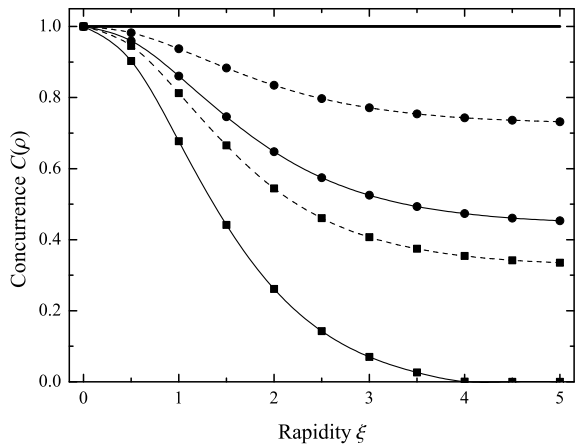}%
\caption{Spin concurrence $C\left(  \rho\right)  $ as a function of rapidity
$\xi$, for an initial Bell state with momentum in an \textquotedblleft
entangled Gaussian\textquotedblright. Data shown as dot (square) is for
$\sigma/m=1$ ($\sigma/m=4$), with solid (dash) line for $x=0$ ($x=0.8$). The
solid line at $C\left(  \rho\right)  =1$ represents the spin concurrence at
the limit $x\rightarrow1$ for any value of $\sigma/m$.}%
\label{Fig1}%
\end{center}
\end{figure}

By integrating over the momenta, we obtain the reduced density matrix, viewed
by the Lorentz-boosted observer, as%
\begin{equation}
\rho=\iint\Psi^{\prime}\left(  \mathbf{p},\mathbf{q}\right)  \Psi^{\prime
}\left(  \mathbf{p},\mathbf{q}\right)  ^{\dag}\widetilde{\text{d}}%
\mathbf{p}\widetilde{\text{d}}\mathbf{q}.
\end{equation}
The entanglement between the spins viewed by the Lorentz-boosted observer is
obtained by calculating the Wootters' concurrence \cite{4}, denoted as
$C\left(  \rho\right)  $. The change in the Lorentz-transformed concurrence
$C\left(  \rho\right)  $ depends on $\sigma/m$, $x$ and $\xi$. Fig. \ref{Fig1}
shows the concurrence as a function of rapidity $\xi$, for different values of
$\sigma/m$ and $x$. Similar to Ref. \cite{2}, the decrease from the maximum
value ($C\left(  \rho\right)  =1$ for Bell states) documents the boost-induced
decoherence of the spin entanglement \cite{2}. However, it is interesting to
see that for fixed $\sigma/m$ and $\xi$, the concurrence decreases less for
non-zero $x$. Further, it is surprising that at the limit $x\rightarrow1$, the
concurrence does not decrease, no matter what $\sigma/m$ and $\xi$ are.
Indeed, at the limit $x\rightarrow1$, not only the concurrence but also the
reduced density matrix for spins are independent of $\sigma/m$ and $\xi$.%

One possible explanation for that the concurrence decreases less with the
presence of momentum entanglement is as follows. Boosting the state, we move
some of the spin entanglement to the momentum \cite{2}, however the momentum
entanglement appears to be moved to spins simultaneously. The transfer of
momentum entanglement to spins hence compensates the decrease of spin
entanglement, and the Lorentz-transformed concurrence decreases less. When the
momenta of the two particles are perfectly correlated, even though may be not
maximally entangled, the transfer of entanglement from momenta to spins
happens to fully compensate the decrease of spin entanglement, so the
entanglement of the reduced spin state remains maximal when viewed by any
Lorentz-boosted observer. Particularly, for the singlet Bell states with
momentum distribution given in Eq. (\ref{eq 4}) (generally in Eq.
(\ref{eq 11}) in the following), the Lorentz boost does not affect the reduced
spin state, only transforms $\mathbf{p}\rightarrow\Lambda\mathbf{p}$ and
$\mathbf{q}\rightarrow\Lambda\mathbf{q}$. The momentum and spin parts of such
states always appear to be separate viewed from any Lorentz-boosted frame.

That the spin concurrence remains maximal at the limit $x\rightarrow1$ when
viewed from any Lorentz-boosted frame can be generalized, without the
assumption that the momentum distribution is an \textquotedblleft entangled
Gaussian\textquotedblright\ given in Eq. (\ref{eq 3}). Directly from Eq.
(\ref{eq 8}), we see that if the momentum distribution takes the following
form,%
\begin{equation}
g^{\prime}\left(  \mathbf{p},\mathbf{q}\right)  =\sqrt{f\left(  \mathbf{p}%
\right)  \delta^{3}\left(  \mathbf{p}-\mathbf{q}\right)  },\label{eq 11}%
\end{equation}
where $f\left(  \mathbf{p}\right)  $ can be any distribution as long as
$g^{\prime}\left(  \mathbf{p},\mathbf{q}\right)  $ is normalized according to
Eq. (\ref{eq 22}), the boosted state could be written as%
\begin{equation}
\Psi^{\prime}\left(  \Lambda\mathbf{p},\Lambda\mathbf{q}\right)
=\frac{g^{\prime}\left(  \mathbf{p},\mathbf{q}\right)  }{\sqrt{2}}\left(
\begin{array}
[c]{c}%
0\\
\alpha_{\mathbf{p}}^{2}+\beta_{\mathbf{p}}^{2}\\
-\alpha_{\mathbf{p}}^{2}-\beta_{\mathbf{p}}^{2}\\
0
\end{array}
\right)  =\Psi\left(  \mathbf{p},\mathbf{q}\right)  ,\label{eq 12}%
\end{equation}
where $\alpha_{\mathbf{p}}^{2}+\beta_{\mathbf{p}}^{2}\equiv1$ due to the
unitarity of $U_{\mathbf{p}}$. For the singlet Bell state shown in Eq.
(\ref{eq 1}) with momentum distribution given in Eq. (\ref{eq 11}), the
reduced density matrix remains the same as in the rest frame when viewed by
any Lorentz-boosted observer. Thus the entanglement between the spins remains
maximal if viewed from any Lorentz-transformed frame. Indeed, the following
four \textquotedblleft Bell\textquotedblright\ states all have invariant
reduced density matrices for spins viewed from any frame Lorentz boosted along
the $z$-axis.
\begin{subequations}
\label{eq 21}%
\begin{align}
\Phi_{f}^{+} &  =\sqrt{f\left(  \mathbf{p}\right)  \delta\left(  p-q\right)
\delta_{\theta_{\mathbf{p}},\theta_{\mathbf{q}}}\delta_{\varphi_{\mathbf{p}%
}+\varphi_{\mathbf{q}},0}}\left\vert \phi^{+}\right\rangle ,\label{eq 21a}\\
\Phi_{f}^{-} &  =\sqrt{f\left(  \mathbf{p}\right)  \delta\left(  p-q\right)
\delta_{\theta_{\mathbf{p}},\theta_{\mathbf{q}}}\delta_{\varphi_{\mathbf{p}%
}+\varphi_{\mathbf{q}},\pi}}\left\vert \phi^{-}\right\rangle ,\label{eq 21b}\\
\Psi_{f}^{+} &  =\sqrt{f\left(  \mathbf{p}\right)  \delta\left(  p-q\right)
\delta_{\theta_{\mathbf{p}},\theta_{\mathbf{q}}}\delta_{\varphi_{\mathbf{p}%
}-\varphi_{\mathbf{q}},\pi}}\left\vert \psi^{+}\right\rangle ,\label{eq 21c}\\
\Psi_{f}^{-} &  =\sqrt{f\left(  \mathbf{p}\right)  \delta^{3}\left(
\mathbf{p}-\mathbf{q}\right)  }\left\vert \psi^{-}\right\rangle
.\label{eq 21d}%
\end{align}
Here we define $\delta_{x,y}\equiv\delta\left(  \left(  x-y\right)
\operatorname{mod}2\pi\right)  $ for compactness of notation. In Eqs.
(\ref{eq 21}), $f\left(  \mathbf{p}\right)  $ could be any distribution as
long as the state is normalized. $\left\vert \phi^{\pm}\right\rangle =\left(
\left\vert \uparrow\uparrow\right\rangle \pm\left\vert \downarrow
\downarrow\right\rangle \right)  /\sqrt{2}$ and $\left\vert \psi^{\pm
}\right\rangle =\left(  \left\vert \uparrow\downarrow\right\rangle
\pm\left\vert \downarrow\uparrow\right\rangle \right)  /\sqrt{2}$ are the
conventional Bell states. Further, the states in Eqs. (\ref{eq 21}), together
with those differing by only rotations, constitute a set of states of which
the entanglement between the spins remains maximal when viewed from any
Lorentz-transformed frame. This invariance of spin entanglement leads to
possible applications to the relativistic quantum information processing. Here
we shall note that, in Eqs. (\ref{eq 21}) as well as in the remaining part of
this paper, the $\delta$-functions should be regard as limits of analytical
functions under certain conditions, such as Eq. (\ref{eq 4}) is the limit of
Eq. (\ref{eq 3}) at $x\rightarrow1$. The only restriction on $f\left(
\mathbf{p}\right)  $ is that the states in Eqs. (\ref{eq 21}) could be normalized.

\section{Relativistic-Invariant Protocol for Quantum Information Processing}

An application of possible interests of the above results is to suggest a
relativistic-invariant protocol for quantum communication. Conventionally
using the spin of a single spin-1/2 particle to represent a qubit may not be
appropriate in relativity theory, because the reduced density matrix for its
spin is generally not covariant under Lorentz transformations \cite{1}. If and
only if for momentum eigenstates (plane waves), the reduced density matrix for
the spin of a single particle could be covariant under Lorentz
transformations, but momentum eigenstates are not localized and difficult for
feasible applications. However, two spin-1/2 particles that are appropriately
entangled, such as in Eqs. (\ref{eq 21}) without being momentum eigenstates,
could indeed have reduced density matrix for spins to be invariant under
Lorentz transformation. Such invariance provides us the possibility to
feasibly represent a single qubit using two appropriately entangled spin-1/2
particles, in a Lorentz-invariant manner. Taking into account that in many
practical situations of communication, one may need to maintain the particles
along desired directions, here we focus on the idea case where the momenta of
the pair of particles have deterministic directions, and assume that the two
particles are moving along the same deterministic direction. We may also
choose the boost $\Lambda$ to be along the $z$-axis, and the momenta to lie in
the $x$-$z$ plane, \textit{i.e.} $\theta_{\mathbf{p}}\equiv\theta_{\mathbf{q}%
}\equiv\theta$ and $\varphi_{\mathbf{p}}\equiv\varphi_{\mathbf{q}}\equiv0$,
without loss of generality. In this protocol we use the momentum distribution
that has the following form in the rest frame,
\end{subequations}
\begin{equation}
\widetilde{g}\left(  \mathbf{p},\mathbf{q}\right)  =\sqrt{f\left(
\mathbf{p}\right)  \delta\left(  p-q\right)  \delta_{\theta_{\mathbf{p}%
},\theta}\delta_{\theta_{\mathbf{q}},\theta}\delta_{\varphi_{\mathbf{p}}%
,0}\delta_{\varphi_{\mathbf{q}},0}},\label{eq 15}%
\end{equation}
with $f\left(  \mathbf{p}\right)  $ being arbitrary as long as $\widetilde
{g}\left(  \mathbf{p},\mathbf{q}\right)  $ is normalized as in Eq.
(\ref{eq 22}). Because Eq. (\ref{eq 15}) is a simultaneous instance of the
momentum distributions of the states in both Eq. (\ref{eq 21a}) and Eq.
(\ref{eq 21d}), both $\widetilde{g}\left(  \mathbf{p},\mathbf{q}\right)
\left\vert \phi^{+}\right\rangle $ and $\widetilde{g}\left(  \mathbf{p}%
,\mathbf{q}\right)  \left\vert \psi^{-}\right\rangle $ have invariant reduced
density matrices for spins when viewed from any Lorentz-boosted frames. This
enables us to use these two states as the orthonormal bases, namely
$\left\vert \widetilde{0}\right\rangle $ and $\left\vert \widetilde
{1}\right\rangle $, of a qubit, as follows.
\begin{subequations}
\label{eq 16}%
\begin{align}
\left\vert \widetilde{0}\right\rangle  & \thicksim\widetilde{g}\left(
\mathbf{p},\mathbf{q}\right)  \left\vert \phi^{+}\right\rangle ,\\
\left\vert \widetilde{1}\right\rangle  & \thicksim\widetilde{g}\left(
\mathbf{p},\mathbf{q}\right)  \left\vert \psi^{-}\right\rangle .
\end{align}
Eqs. (\ref{eq 16}) could be regarded as a representation of a
\textquotedblleft Lorentz-invariant\textquotedblright\ qubit, in the sense
that we look only at the spin part of the state. The representation of
\textquotedblleft Lorentz-invariant\textquotedblright\ multiple qubits could
be obtained straightforward. Note that in multi-qubit states, the momentum
distributions of individual qubits are not necessarily the same. We can
further find operator acting upon a single qubit, in terms of the
\textquotedblleft Lorentz-invariant\textquotedblright\ bases, as
\end{subequations}
\begin{equation}
\widetilde{\mathcal{O}}=\sum\nolimits_{\sigma,\tau=0,1}\lambda_{\sigma\tau
}\left\vert \widetilde{\sigma}\right\rangle \left\langle \widetilde{\tau
}\right\vert .\label{eq 24}%
\end{equation}
The operators acting upon multiple qubits can be obtained analogously. We
refer the operators as in Eq. (\ref{eq 24}) to be \textquotedblleft
Lorentz-invariant\textquotedblright\ in the sense that, if we look only at
spins, the action of the operator on the state $a\left\vert \widetilde
{0}\right\rangle +b\left\vert \widetilde{1}\right\rangle $ ($\forall
a,b\in\mathbb{C}$ with $\left\vert a\right\vert ^{2}+\left\vert b\right\vert
^{2}=1$) remains the same when viewed in any Lorentz-boosted frame. Within the
set of these \textquotedblleft Lorentz-invariant\textquotedblright\ qubits and
operators, the entropy, entanglement and measurement results all have
invariant meanings, despite that for a single quantum spin and some other
situations these quantities may have no invariant meanings in different frames
\cite{1,2}. Therefore it is guaranteed that, using such states and operators,
the non-relativistic quantum information theory could be invariantly applied
to relativistic situation.

\section{Conclusion}

As observed in Ref. \cite{2}, because Lorentz boosts entangle the spin and
momentum degrees of freedom, entanglement between the spins may change if
viewed from a moving frame. Especially, maximally entangled spin states will
most likely decohere due to the mixing with momentum degrees of freedom,
depending on the initial momentum wave function \cite{2}. In this paper, we
investigate the quantum entanglement between the spins of a pair of spin-1/2
massive particles in moving frames, for the case that the momenta of the
particles are entangled. We show that if the momenta of the pair are
appropriately entangled, the entanglement between the spins of the Bell states
remains maximal when viewed from any Lorentz-transformed frame. Further, we
suggest a relativistic-invariant protocol for quantum communication, with
which the non-relativistic quantum information theory could be invariantly
applied to relativistic situations. Though the investigations are based on
spin-1/2 particles, we believe the similar results for larger spins could be
obtained analogously. Especially, we hope our work would help to find a
relativistic-invariant protocol for quantum information processing based on
photons, \textit{i.e.} the case of massless spin-1 particles.

\begin{acknowledgments}
We would like to thank M.J. Shi, Z.B. Chen, and Y.S. Xia for the fruitful
discussions. We also acknowledge the kind help on the subject and valuable
suggestions from R.M. Gingrich, D.R. Terno, and A. Peres. This work was
supported by the Nature Science Foundation of China (Grant No. 10075041), the
National Fundamental Research Program (Grant No. 2001CB309300), and the ASTAR
Grant No. 012-104-0040.
\end{acknowledgments}

\end{document}